\documentstyle[12pt,epsfig]{article}
\def\oot{\frac{1}{2}}

\begin{document}
\begin{titlepage}

\vspace{1cm}

\begin{center}

{\Large \bf   Gravitational-Wave Stochastic Background from  Kinks
and Cusps on Cosmic Strings } \vspace{0.5cm} \vspace{0.5cm}
\vspace{0.5cm}

\end{center}

\begin{center}
{{\bf S. \"Olmez}$^{\,a}$, {\bf V.~Mandic}$^{\,a}$ and {\bf
X.~Siemens}$^{\,b}$  }
\end {center}
$^a${\it Department of Physics and Astronomy, University of
Minnesota, Minneapolis, MN 55455, USA}\\
$^b${\it  Center for Gravitation and Cosmology, Department of
Physics, P.O. Box 413, University of
Wisconsin - Milwaukee, Wisconsin, 53201, USA}\\

\begin{center}
{\bf Abstract}
\end{center}
We compute the contribution of kinks on cosmic string loops to
stochastic background of gravitational waves (SBGW). We find that
kinks contribute at the same order as cusps to the SBGW. We discuss
the accessibility of the total background due to kinks as well  as
cusps to current and planned gravitational wave detectors, as well
as to the big bang nucleosynthesis (BBN), the cosmic microwave
background (CMB), and pulsar timing constraints. As in the case of
cusps, we find that current data from interferometric gravitational
wave detectors, such as LIGO, are sensitive to areas of parameter
space of cosmic string models complementary to those accessible to
pulsar, BBN, and CMB bounds.
\end{titlepage}

\bibliographystyle{plain}

\pagestyle{plain}
\section{Introduction}
Topological defects are remnants of spontaneously broken local or
global symmetries. The simplest and the most well-known example of
the former one is the Abrikosov-Nielsen-Olesen flux tube \cite{ANO},
which originates from spontaneously broken $U(1)$ gauge symmetry.
Most of the attention in the literature has been focused on defects
originating from broken gauge symmetries, since grand unified
theories have gauge symmetries which are eventually spontaneously
broken down to the symmetry of the Standard Model. Cosmic strings
are one dimensional topological defects predicted by a large class
of unified theories \cite{kibble76,alexbook,Jeannerot:2003qv}.
Cosmic strings were first considered as the seeds of structure
formation \cite{Zeldovich,AV81a}, however, later, it was discovered
that cosmic strings were incompatible with the cosmic microwave
background (CMB) angular power spectrum. Cosmic strings can still
contribute to structure formation, but they cannot be the dominant
source.  Cosmic strings are also candidates for the generation of
other observable astrophysical phenomena such as high energy cosmic
rays, gamma ray burst and gravitational waves \cite{alexbook, book
anderson,khlopov1,khlopov2}. Furthermore, recently it has been shown
that in string-theory-inspired cosmological scenarios cosmic strings
may also be generated~\cite{cstrings}. They are referred to as
cosmic superstrings. This realization has revitalized interest in
cosmic strings and their potential observational signatures. There
are some important differences between cosmic strings and cosmic
superstrings. The reconnection probability is unity for cosmic
strings \cite{alexbook,Nitta}. Cosmic superstrings, on the other
hand, have reconnection probability less than unity. This is a
result of the probabilistic nature of their interaction and also the
fact that it is less probable for strings to meet since they can
live in higher dimensions~\cite{dvali vilenkin 2004}. The value of
$p$ ranges from $10^{-3}$ to $1$ in different theories \cite{jjp}.
Cosmic superstrings could also be unstable, decaying long before the
present time. In this case, however, they may also leave behind a
detectable gravitational wave signature~\cite{Leblond:2009fq}.

In the early universe, a network of cosmic strings evolves toward to
an attractor solution called  the ``scaling regime". In the scaling
regime the statistical properties of the network, such as the
average distance between strings and the size of loops at formation,
scale with the cosmic time. In addition, the energy density of the
network remains a small constant fraction of the energy density of
the universe. For cosmic superstrings in the scaling regime, the
density of the network $\rho$ is inversely proportional to the
reconnection probability $p$, that is $\rho\propto p^{-\beta}$. The
value of $\beta$ is still under debate~\cite{stoica
tye,Sakellariadou 2005,avgoustidis shellard}, and as a placeholder
in our analysis we assume that $\beta=1$.

The gravitational interaction of strings is characterized by their
tension $\mu$, or more conveniently by the dimensionless parameter
$G \mu$, where $G$ is Newton's constant. The current CMB bound on
the tension is $G \mu<6.1\times 10^{-7}$ \cite{CMB1,CMB2}. It was
first believed that gravitational radiation from cosmic strings with
$G \mu\ll 10^7$ would be too weak to observe. However it was later
shown that gravitational radiation produced at cusps, which have
large Lorentz boosts, could lead to a detectable signal~\cite{DV1,DV2,SCMMCR}. Gravitational
radiation bursts from (super)strings could be observable by current
and planned gravitational wave detectors for values of $G \mu$ as
low as $10^{-13}$, which may provide a test for a certain class of string theories \cite{pol1}.
Indeed, searches for burst signals using ground-based detectors are already underway~\cite{Abbott:2009rr}.

A gravitational background produced by the incoherent superposition
of cusp  bursts from a network of cosmic strings and superstrings
was considered in \cite{Siemens Mandic Creighton}. In this paper we
extend this computation to include kinks, long-lived sharp edges on
strings that result from intercommutations, and find that kinks
contribute at almost the same level as cusps. We investigate the
detectability of the total background produced by cusps and kinks by
a wide range of current and planned experiments. A similar
calculation for the case of infinite strings has been undertaken in
the recent paper \cite{Kawasaki}, see also \cite{Kawasaki2}.

 The organization of the paper is as follows: In Sect.
\ref{Gravitational Radiation} we consider gravitational waves
generated by cusps and kinks in the weak field limit
\cite{Weinberg}. In this section we follow the conventions of
\cite{DV1,DV2}, and more details can be found in these references.
In Sect. \ref{Stochastic Background} we derive the expression for
the stochastic background, which is a double integral over redshift
and loop length. In Sect. \ref{Analytical app} we evaluate integral
analytically with certain approximations, which results in a flat
distribution for larger values of the frequency. Finally in Sec.
\ref{parameter scan} we numerically evaluate the background and
discuss the observability by various experiment.
\newpage
\section{Gravitational Radiation}\label{Gravitational Radiation}
In this section we  consider gravitational waves created by cusps
and kinks. For completeness we follow closely the analysis in \cite{DV1,DV2}, and reproduce a number of their results.
We begin with a derivation for the metric pertubation in terms of the Fourier transform
of the stress energy tensor of the source. We then write the stress energy tensor for a relativistic string
and compute its Fourier transform. Using these results we then compute the gravitational waveforms
produced by cusps and kinks on cosmic strings.

\subsection{Calculation of metric perturbations}\label{Calculation of metric
perturbations}
 \flushleft Gravitational waves from a source can be
calculated using the weak field approximation \cite{Weinberg},
\begin{equation}\label{weak field expansion}
g_{\mu\nu}=\eta_{\mu\nu}+h_{\mu\nu},
\end{equation}
where $\eta_{\mu\nu}$ is the Minkowski metric with positive
signature and $h_{\mu\nu}$ is the metric perturbation. In the
harmonic gauge, $g^{\mu\nu}\Gamma^\lambda_{\mu\nu}=0$, the
linearized Ricci tensor is
\begin{equation}\label{weak field expansion of ricci in harmonic gauge}
R_{\mu\kappa}\simeq \oot \partial_\lambda\partial^\lambda
h_{\mu\kappa}.
\end{equation}
Substituting into  Einstein's equations yields
\begin{equation}\label{weak field expansion of Einstein Equation in harmonic gauge 1}
R_{\mu\nu}-\oot g_{\mu\nu} R\simeq
\oot(\partial_\lambda\partial^\lambda h_{\mu\nu}-\oot
\eta_{\mu\nu}\partial_\lambda\partial^\lambda h)=-8\pi G {\mathcal
T}_{\mu\nu},
\end{equation}
where $R$ is the Ricci scalar, ${\mathcal T}_{\mu\nu}$ is the energy
momentum tensor of matter and $h=\eta_{\mu\nu}h^{\mu\nu}$. Defining
$\bar{h}_{\mu\nu}= h_{\mu\nu}-\oot \eta_{\mu\nu}h$ further
simplifies Eq. (\ref{weak field expansion of Einstein Equation in
harmonic gauge 1}),
\begin{equation}\label{weak field expansion of Einstein Equation in harmonic gauge 2}
\partial_\lambda\partial^\lambda \bar{h}_{\mu\nu}=-16\pi G
{\mathcal T}_{\mu\nu},
\end{equation}
which is a wave equation with a source term. We can rewrite this
equation in the frequency domain as,
\begin{equation}\label{green's function equation}
(w^2+\nabla^2)\bar{h}_{\mu\nu}(\vec{x},w)=-16\pi G {\mathcal
T}_{\mu\nu}(x,w),
 \end{equation}
 where
\begin{equation}\label{weak field expansion of Einstein Equation in harmonic gauge}
 \bar{h}_{\mu\nu}(\vec{x},w)=\int dt\,e^{i w
 t}\bar{h}_{\mu\nu}(\vec{x},t).
 \end{equation}
Eq. (\ref{green's function equation}) can be solved by using the
Green's function for the operator $w^2+\nabla^2$, which is
\begin{equation}\label{green's function solution}
{\mathcal G}(\vec{x}-\vec{x}',w)=\frac{e^{i w
|\vec{x}-\vec{x}'|}}{|\vec{x}-\vec{x}'|}.
 \end{equation}
Therefore metric perturbations are given by
  \begin{eqnarray}\label{metric perturbations final}
\bar{h}_{\mu\nu}(\vec{x},w)&=& -16 \pi G \int d^3x'{\mathcal G}(\vec{x}-\vec{x}',w){\mathcal T}_{\mu\nu}(\vec{x}',w)\nonumber\\
&=&-16 \pi G \frac{e^{i w |\vec{x}|}}{|\vec{x}|} {\mathcal
T}_{\mu\nu}(\vec{k},w),
 \end{eqnarray}
 where $\vec{k}=w \hat{x}$ and
  \begin{eqnarray}\label{EMT fourier transform}
{\mathcal T}_{\mu\nu}(\vec{k},w)&=&\frac{1}{T}\int_0^T dt\int
d^3x'e^{i(wt- \vec{k}\cdot\vec{x}')}{\mathcal
T}_{\mu\nu}(\vec{x}',t),
 \end{eqnarray}
 where $T$ is the fundamental period of the source. Eq. (\ref{metric perturbations
 final}) relates energy momentum tensor to gravitational waves. The
 next step is to calculate the energy momentum tensor of cusps and
 kinks on cosmic strings.
 \subsection{Energy Momentum Tensor of Cosmic Strings} $\newline$
In the thin wire approximation, the dynamics of strings is described
by the Nambu-Goto action \cite{alexbook, book anderson}
  \begin{eqnarray}\label{Nambu-Goto action}
S=-\mu \int d\tau d\sigma \sqrt{-\gamma},
 \end{eqnarray}
where $\sigma$ and $\tau$ are world-sheet coordinates and $\mu$ is
the string tension. $\gamma$ is the determinant of the induced
metric
\begin{equation}\label{induced metric}
   \gamma_{a\,b}=\eta_{\mu\nu}\partial_a X^\mu\partial_b X^\nu,
\end{equation}
where $a$ and $b$ denote world sheet coordinates. The  equation of
motion following from Eq. (\ref{Nambu-Goto action}) is
 \begin{eqnarray}\label{Eqn of motion}
&&(\partial_\tau^2-\partial_\sigma^2)X^\mu=0,
 \end{eqnarray}
The solution must also satisfy Virasoro conditions
  \begin{eqnarray}\label{virasoro}
&&\dot X\cdot\dot X+X'\cdot X'=0\;\; \rm{and} \;\;\ \dot X\cdot
X'=0,
 \end{eqnarray}
 where  $dot$ and $prime$ denote derivatives with respect to $\tau$ and $\sigma$ respectively.
 If we define $\sigma_\pm= \tau\pm\sigma$, the equation of motion
 becomes
  \begin{eqnarray}\label{Eqn of motion light cone}
\partial_+\partial_-X^\mu&=&0,
 \end{eqnarray}
which is solved by left and right moving waves,
  \begin{eqnarray}\label{left right}
X^\mu&=&\oot \left(X^\mu_+(\sigma_+)+X^\mu_-(\sigma_-)\right).
 \end{eqnarray}
 Furthermore Virasoro conditions in Eq. (\ref{virasoro}) simplify to
   \begin{eqnarray}\label{virasoro 2}
&&{\bf\dot X_\pm}^2=1,
 \end{eqnarray}
  where $dot$ now represents the derivative with respect to the (unique) argument
of the functions $X^\mu_\pm$.
 We require that $X^\mu(\sigma,\tau)$ is periodic in $\sigma$ with period $l$, which is the length of the loop. This implies that the functions
 $X^\mu_\pm$ are periodic functions with the same period. The
 period in $t$ is $l/2$ since
 $X^\mu(\sigma+l/2,\tau+l/2)=X^\mu(\sigma,\tau)$.\\
 The  energy momentum tensor corresponding to the Nambu-Goto action can be calculated by varying
Eq. (\ref{Nambu-Goto action}) with respect to the metric, which
yields
 \begin{eqnarray}\label{EMT of Nambu-Goto}
{\mathcal T}_{\mu\nu}(x)&=&-2 \frac{\delta S}{\delta
\eta_{\mu\nu}}=\mu\int d\tau d\sigma (\dot X^\mu \dot X^\nu- X'^\nu
          X'^\mu)\,\delta^{(4)}(x-X)\nonumber\\
          &=&\frac{\mu}{2}\int d\sigma_-d\sigma_+
 (\dot X_+^\mu \dot X_-^\nu+ \dot X_-^\nu
          \dot X_-^\mu)\,\delta^{(4)}(x-X).
 \end{eqnarray}
 Inserting this expansion into Eq. (\ref{EMT fourier transform}) gives us the energy momentum tensor in momentum space
  \begin{eqnarray}\label{EMT left right}
{\mathcal T}_{\mu\nu}(k)&=&\frac{\mu}{T_l}\int d\sigma_-d\sigma_+
\dot X_+^{(\mu} \dot X_-^{\nu)}e^{-\frac{i}{2}(k\cdot X_++k\cdot
X_-)},
 \end{eqnarray}
where we define
  \begin{eqnarray}\label{symmetrization}
 \dot X_+^{(\mu} \dot
X_-^{\nu)}&=&\oot(\dot X_+^{\mu} \dot X_-^{\nu}+\dot X_-^{\mu} \dot
X_+^{\nu}).
 \end{eqnarray}
 The nice property of Eq. (\ref{EMT left right}) is that two
 integrals can be calculated independently,
   \begin{eqnarray}\label{I}
I_\pm^\mu(k)&\equiv&\int_0^l d\sigma_\pm \dot X_\pm^{\mu}
e^{-\frac{i}{2}k\cdot X_\pm},
 \end{eqnarray}
and the energy momentum tensor can be expressed in terms of
$I_\pm^\mu$ as
 follows;
   \begin{eqnarray}\label{EMT left right final}
{\mathcal T}_{\mu\nu}(k)&=&\frac{\mu}{l}I_+^{(\mu} I_-^{\nu)},
 \end{eqnarray}
where we used $T_l=\frac{l}{2}$.  In the following subsection we
calculate $I_\pm^\mu$ for cusps and kinks.
\subsubsection{Cusps} $\newline$
Let us start with the geometrical interpretation of Eq.
(\ref{virasoro 2}). It tells us that ${\bf\dot X_\pm}$ trace a unit
sphere centered at the origin, which is called Kibble-Turok sphere.
Integrating ${\bf\dot X_\pm}$ and using the periodicity, we get
\begin{equation}\label{turok}
    \int_0^l{\bf\dot X_\pm}(\sigma_\pm)d\sigma_\pm=0,
\end{equation}
which implies that ${\bf\dot X_\pm}$ cannot lie completely in a
single hemisphere and therefore they intersect at some point(s). We
choose our parametrization and the coordinate system such that the
intersection occurs at the parameters $\sigma_\pm=0$ at the origin,
that is $X_\pm^\mu(0)=0$. $X_\pm(\sigma_\pm)$  and $\dot
X_\pm(\sigma_\pm)$ can be expanded around $\sigma_\pm=0$
\begin{eqnarray}
X_\pm^\mu(\sigma_\pm)&=&l_\pm^\mu \sigma_\pm+\oot\ddot X_\pm^\mu
\sigma_\pm^2 +\frac{1}{6}X_\pm^{(3)\mu}\sigma_\pm^3\label{cusp
expansion} \\
\dot X_\pm^\mu(\sigma_\pm) &=&l_\pm^\mu +\ddot X_\pm^\mu \sigma_\pm
+\frac{1}{2}X_\pm^{(3)\mu}\sigma_\pm^2.\label{cusp derivative
expansion}
\end{eqnarray}
where $l_\pm^\mu=\dot X_\pm^\mu(0)$. We can easily find the shape of
$ X_\pm^\mu$ at $\tau=0$ ($\sigma_\pm=\pm\sigma$),
\begin{eqnarray}\label{cusp shape}
X^\mu(\sigma,\tau=0)&=&\frac{1}{2}\left(X_+^\mu(\sigma)+X_-^\mu(-\sigma)\right)\nonumber\\
&=&\frac{1}{4}(\ddot X_+^\mu+\ddot X_-^\mu)\sigma^2
+\frac{1}{12}(X_+^{(3)\mu}+X_-^{(3)\mu})\sigma^3.
\end{eqnarray}
In order to visualize the shape of the string around the origin, we
can choose the coordinate system such that $(\ddot{\vec{X}_+}+\ddot{
\vec{X}_-})$ lies on the $x$-axis, and define
$x=\frac{1}{4}|\ddot{\vec{X}_+}+\ddot{ \vec{X}_-}|\sigma^2$. Let us
also denote the direction of $\vec{X}_+^{(3)}+\vec{X}_-^{(3)}$ by
$\hat{y}$, which is not necessarily orthogonal to $\hat{x}.$ If we
define $y=\frac{1}{12}|X_+^{(3)\mu}+X_-^{(3)\mu}|\sigma^3$, we see
that $y\propto x^{\frac{3}{2}}$, which has a sharp turn at $x=0$, which is referred to as cusp.\\
 We can calculate $I_\pm^\mu$ for cusps using the expansion in Eq. (\ref{cusp expansion}). First of all, we  note that the first
term Eq. (\ref{cusp derivative expansion}) is pure gauge, it can be
removed by a coordinate transformation. Furthermore imposing
Virasoro condition in Eq. (\ref{virasoro 2}) gives
\begin{equation}\label{cusp expansion2}
 l_\pm\cdot\ddot X_\pm=0,\;\;\;\rm{and}\;\;\; l_\pm\cdot X^{(3)}_\pm=-\ddot
 X^2_\pm.
\end{equation}
 When the line of sight $k$ is in the direction of $l$ we have $k= w l$, which gives
 $-i\,k\cdot X_\pm = \frac{i}{6}w\ddot X_\pm^2
 \sigma_\pm^3$.
If we plug in the expansion in Eq. (\ref{cusp expansion}) into  Eq.
(\ref{I}) we get,
    \begin{eqnarray}\label{EMT left or right expanded}
I_\pm^\mu(k)&=&\ddot X_\pm^\mu\int_0^l d\sigma\, \sigma
e^{\frac{i}{12}  w \ddot X_\pm^2 \sigma^3}=\frac{2 \pi i \ddot
X_\pm^\mu}{3 \Gamma(1/3)\left(\frac{1}{12}  w |\ddot
X_\pm^2|\right)^{2/3}} .
 \end{eqnarray}
 Replacing $w$ with $2 \pi f$ gives
    \begin{eqnarray}\label{Imu final}
I_\pm^\mu(k)&=&C_\pm^\mu f^{-\frac{2}{3}},\\
{\mathcal T}_{\mu\nu}(k)&=&\frac{\mu}{l}|f|^{-\frac{4}{3}}
C_+^{(\mu} C_-^{\nu)}\label{emt final}
 \end{eqnarray}
where $C_\pm^\mu= i \frac{(32 \pi/3)^{1/3}}{\Gamma(1/3)} \frac{\ddot
X_\pm^\mu}{|\ddot X_\pm|^{\frac{4}{3}}} $. Finally we need to
estimate $|\ddot X_\pm|=|{\bf\ddot X_\pm}|$. Since ${\bf X}_\pm$ is
periodic with period $l$, ${\bf \dot X}$ expanded as
\begin{equation}\label{X dot expansion}
{\bf \dot X}(\sigma_\pm)=\sum_n {\bf c_n} e^{i \frac{2 \pi}{l}n
\sigma_\pm},
\end{equation}
where the expansion coefficients ${\bf c_n}$ are constrained by
$|\dot {\bf X}_\pm|=1$. If the string is not too wiggly, ${\bf c_n}$
is nonvanishing for only small $n$, therefore we can estimate
$|\ddot X_\pm|\sim \frac{2 \pi}{l}$. Combining all the pieces
together and neglecting decimal points in the numerical coefficient,
we express the trace of the metric perturbations as\
    \begin{eqnarray}\label{cusp radiation final}
h^{(c)}(f)\equiv |\bar h^\mu_\mu|= \frac{G \mu l^{\frac{2}{3}}
}{r}|f|^{-\frac{4}{3}}.
 \end{eqnarray}
We can express $r$ as a function of $z$
 \begin{equation}\label{r redshift}
    r=\frac{1}{H_0}\int_0^z\frac{dz'}{{\mathcal H}(z')}\equiv \frac{1}{H_0}\varphi_r(z),
 \end{equation}
where $H_0$ is the Hubble constant today and ${\mathcal H}(z)$ is
the Hubble function given by
 \begin{equation}\label{hubble}
    {\mathcal H}(z)=\left(\Omega_M(1+z)^3+\Omega_R(1+z)^4+\Omega_\Lambda\right)^{1/2}.
\end{equation}
The numerical values for the constants in this equation are
$\Omega_M=0.25$, $\Omega_R=4.6\times 10^{-5}$,
$\Omega_\Lambda=1-\Omega_R-\Omega_M$ and $H_0= 73 \rm{km/s/Mpc}$.\\
 Note that $f$ in Eq. (\ref{cusp radiation final}) is the frequency of the
radiation  in the frame of emission. In order to convert it to the
frequency we observe today, the effect of the cosmological redshift
must be included. The frequency in the frame of emission, $f$, is
related to the frequency we observe now, $f_{\rm{now}}$, by the
relation $f=(1+z)f_{\rm{now}}$. After redshifting properly\footnote{
One should note that replacing $f$ in Eq. (\ref{cusp radiation
final}) with $(1+z)f_{\rm{now}}$ is {\it not} correct, since this
replacement will scale the argument and the amplitude of
$h^{(c)}(f)$ by a factor of $\frac{1}{1+z}$, which is the reflection
of the fact that the measure of Fourier integral is not
dimensionless. Since redshifting should change the argument but not
the amplitude, one needs to multiply the result by $1+z$ so that the
amplitude remains the same. Equivalently, one can define Logarithmic
Fourier Transform, as discussed in Ref. \cite{DV1}, such that the
measure of the transform becomes dimensionless.}, Eq. (\ref{cusp
radiation final}) becomes
\begin{eqnarray}\label{cusp radiation in redshift}
h^{(c)}(f,z,l)= \frac{G \mu H_0\,l^{\frac{2}{3}}
}{(1+z)^{\frac{1}{3}}\varphi_r(z)}|f|^{-\frac{4}{3}},
 \end{eqnarray}
where we dropped the subscript now.
\subsubsection{ Kinks} $\newline$
Calculation of kink radiation is similar to the cusp case. The form
of $I_+^\mu$ is the same as the cusp result. $I_-^\mu$ has a
discontinuity at the cusp point and needs a different treatment. Let
us describe the kink (at $\sigma_-=0$ and $X_\pm=0$) as a jump of
the tangent vector from $l_1^\mu$ to $l_2^\mu$. At the first order
one can replace  approximate $\dot X_-^\mu$  by $l_1^\mu$ for
$\sigma_-<0$ and $l_2^\mu$ for $\sigma_->0$. At this approximation,
one gets
    \begin{eqnarray}\label{I- for kink}
I_-^\mu(k)&=&\int_{-l/2}^{l/2} d\sigma_-\dot X_-^\mu e^{-\frac{i}{2}
k \cdot X_-} \simeq \frac{2 i}{w}\left( \frac{l_1^\mu}{l_1\cdot \hat
k}-\frac{l_2^\mu}{l_2\cdot \hat k}\right),
 \end{eqnarray}
 where we dropped two oscillatory terms.
 The exact value of Eq. (\ref{I- for kink}) depends on the sharpness
 of the kink, $l_1\cdot l_2$  \cite{copeland kibble}, however we will assume that
 the average value of this quantity is of order one.
Combining this result with $I_+^\mu$ we get the frequency
distribution of the radiation from a kink as
    \begin{eqnarray}\label{kink radiation redhift}
h^{(K)}(f,z,l)= \frac{G \mu l^{\frac{1}{3}}
H_0}{(1+z)^{\frac{2}{3}}\varphi_r(z)}f^{-5 /3}.
 \end{eqnarray}
 It is important to note that in the derivation of  Eqs. (\ref{cusp radiation in redshift})
and (\ref{kink radiation redhift}) we assumed that the line of sight
$k^\mu$ is in the direction of the motion of the cusp or kink,
$l^\mu$. It is easy to show that $I_\pm$ (Eq. (\ref{I})) decay
exponentially with the angle between ${\bf k}$ and ${\bf l}
$\cite{DV2} . Therefore Eqs. (\ref{cusp radiation in redshift}) and
(\ref{kink radiation redhift}) are valid for angles smaller than
\begin{equation}\label{theta max}
    \theta_m=\frac{1}{\left(f l (1+z)\right)^{\frac{1}{3}}}.
\end{equation}
We  implement this condition with a $\Theta$-function in the
amplitude.
\section{Stochastic Background}\label{Stochastic Background}
The stochastic gravitational background \cite{Siemens Mandic
Creighton} is given by
\begin{eqnarray}\label{background}
    \Omega_{gw}(f)&=&\frac{4\pi^2}{3 H_0^2}f^3\int dz\int dl\,
    h^2(f,z,l)\frac{d^2R(z,l)}{dz dl},
    \end{eqnarray}
where $h(f,z,l)$ is given in Eqs. (\ref{cusp radiation in redshift})
and (\ref{kink radiation redhift}) and $\frac{d^2R(z,l)}{dz dl}$ is
the observable burst rate per length per redshift, which will be
defined below. We take the number of cusps (kinks) to be one per
loop. If we define the density (per volume) of the loops of length
$l$ at time $t$ as $n(l,t)$, the rate of burst (per loop length per
volume) can be expressed as $\frac{ n(l,t)}{l/2}$, where $l/2$
factor is the fundamental period of the string. However, this is not
the observable burst rate since we can observe only the fraction of
bursts that is beamed toward us. Including this fraction we obtain
\begin{equation}\label{rate0}
    \frac{dR}{dl dz}=H_0^{-3}\varphi_V(z)(1+z)^{-1}
    \frac{2 n(l,t)}{l}\Delta(z,f,l),
\end{equation}
where  $(1+z)^{-1}$ comes from converting emission rate to observed
rate, and $H_0^{-3}\varphi_V(z)$ follows from converting
differential volume element to the corresponding function of
redshift $z$,
\begin{equation}\label{differential volume}
    dV=4\pi a^3(t) r^2 dr=\frac{4\pi H_0^{-3}\varphi^2_r(z)}{(1+z)^3 {\mathcal H}(z)}dz
\equiv H_0^{-3} \varphi_V(z) dz,
\end{equation}
where $a(t)$ is the cosmological scale factor.  $\Delta(z,f,l)$ is
the fraction of the bursts we can observe. Geometrically the
radiation from a cusp will be in a conic region with half opening
angle $\theta_m$ (Eq. (\ref{theta max})) and outside the cone it
will decay exponentially. To simplify the calculation we assume that
the radiation amplitude vanishes outside this conic region, which
will be implemented by a $\Theta$-function. We can express the
corresponding  solid angle in terms of the opening angle by using
the following relation
\begin{equation}\label{angle}
 \Omega_m=2 \pi (1-\cos\theta_m)\simeq \pi \theta_m^2.
\end{equation}
Thus the probability that the line of sight is within this solid
angle is
\begin{equation}\label{beaming factor}
 \frac{\Omega_m}{4 \pi}\simeq  \theta_m^2/4,
\end{equation}
which is referred to as the beaming fraction of the cusp.
We combine the cutoff for large angles and beaming effect into
\begin{equation}\label{delta}
    \Delta(z,f,l)\approx\frac{\theta_m^2(z,f,l)}{4}
    \Theta({1-\theta_m(z,f,l)}).
\end{equation}
It is important to note that cusps are instantaneous events, and it is possible
to observe their radiation only if the line of sight happens to be inside
the cone of radiation. The beaming fraction, Eq.
(\ref{beaming factor}), which is proportional to $\theta_m^2$, is the fraction of the time the line of sight is inside the cone of radiation. In contrast,
kinks radiate continuously--as kinks travel around a string loop they radiate in a fan-like pattern.
Therefore radiation cone of a kink will
sweep a strip of width $2 \theta_m$ and an average length $\pi$ on
the surface of the unit sphere as it travels around the cosmic string loop. That is, the probability of
observing radiation from a kink is
\begin{equation}\label{angle prob kink}
 \frac{\Omega^c_m}{4 \pi}\simeq \frac{2 \theta_m \pi}{4 \pi}=\frac{ \theta_m
 }{2}.
\end{equation}
For kinks the cutoff for large angles and beaming factor that enters the rate is therefore
\begin{equation}\label{delta kink}
    \Delta^{(K)}(z,f,l)\approx\frac{\theta_m(z,f,l)}{2}
    \Theta({1-\theta_m(z,f,l)}).
\end{equation}
Inserting this result into Eq. (\ref{background}) gives the
background radiation $\Omega_{gw}(f)$ as a double integral over
$l$ and $z$, which needs to be evaluated numerically. Finally we
need to discuss the form of the loop density, $n(l,t)$ in Eq. (\ref{rate0}).
To do this, it is convenient to first convert the cosmic time $t$ to a suitable
function of redshift $z$ using the following relation
\begin{equation}\label{time to z}
    \frac{dz}{dt}=-(1+z)H_0 \,{\mathcal H}(z),
\end{equation}
which can be integrated to give
\begin{equation}\label{time}
    t=H_0^{-1} \int_z^\infty \frac{dz'}{(1+z'){\mathcal H}(z')}= H_0^{-1}
    \varphi_t(z).
\end{equation}
Below we discuss the two main contending scenarios for the size
of cosmic string loops.

\subsection{Small Loops}\label{small loops}
Early simulations suggested that the size of loops was dictated by
gravitational back reaction. In this case the size of the loops is
fixed by the cosmic time $t$, and all the loops present at a cosmic
time $t$, are of the same size $\alpha\, t$. The value of $\alpha$
is set by the gravitational back reaction, that is $\alpha \propto
\Gamma G \mu$ (In Sect. \ref{parameter scan} we parameterize
$\alpha$ by $\alpha =\epsilon\Gamma G \mu$ where $\epsilon$ is a
parameter we scan over.) The constant $\Gamma$ is the ratio of the
power radiated into gravitational waves by loops to $G\mu^2$.
Numerical simulation results suggest that $\Gamma \sim 50$.
Therefore the density is of the form
\begin{equation}
n(l,t) \propto(p \,\Gamma G \mu)^{-1} t^{-3} \delta(l-\alpha t),
\label{density small loop}
\end{equation}
where $p$ is the reconnection probability. The overall coefficient
is estimated by simulations (for a review see~\cite{alexbook}) which show
that the density in the radiation domination era is about $10$ times
larger the one in the matter domination era. This behavior of the
density can be implemented by a function, $c(z)$, which converges to
$10$ for $z\gg z_{eq}$ and to $1$ for $z\ll z_{eq}$. Therefore the
density can be written as
\begin{equation}
n(l,t) = c(z)(p \,\Gamma G \mu)^{-1} t^{-3} \delta(l-\alpha t),
\label{density small loop 2}
\end{equation}
where~\cite{DV1}
\begin{equation}
c(z)=1+\frac{9 z}{z+z_{eq}}.
 \label{c of z}
\end{equation}
Such a distribution simplifies the calculation of SBGW since the
$l$-integral in Eq. (\ref{background}) can be evaluated trivially to
yield
\begin{eqnarray}\label{z small loop cusp}
    \Omega_{gw}(f)&=&\frac{4\pi^2}{3 H_0^2}\int dz\int dl\,
    h^2(f,z,l)\frac{d^2R(z,l)}{dz dl}\nonumber\\
    &=&\frac{2 \,c \,G\mu\, \pi^2
    H_0^{1/3}}{3 \,p\,\alpha^{1/3}\Gamma f^{1/3}}\int dz\frac{c(z)\varphi_V
    \Theta\left(1-\left[f(1+z)\alpha \varphi_t\right]^{-1/3}\right) }
    {(1+z)^{7/3}\varphi_r^2\varphi_t^{10/3}
    }.
\end{eqnarray}
For kinks, we have a similar integral ,
\begin{eqnarray}\label{z small loop kink}
    \Omega^K_{gw}(f)&=&\frac{4 \,c \,G\mu\, \pi^2
    H_0^{1/3}}{3 \,p\,\alpha^{2/3}\Gamma f^{2/3}}\int dz\frac{c(z)\varphi_V
    \Theta\left(1-\left[f(1+z)\alpha \varphi_t\right]^{-1/3}\right) }
    {(1+z)^{8/3}\varphi_r^2\varphi_t^{11/3}
    }.
\end{eqnarray}
We analytically evaluate the integrals in Eqs. (\ref{z small loop
cusp}) and (\ref{z small loop kink}) in Sec. \ref{Analytical app}
with certain approximations, and perform numerical integration in
Sec. \ref{parameter scan}.

\subsection{Large Loops}\label{large loops}
 Recent simulations \cite{RSB,shellardrecentsim,VOV} suggest
that the size of the loops is set by the large scale dynamics of the
network, and that the gravitational back-reaction scale is irrelevant. In
Ref. \cite {VOV} is found that the loop production functions have
peaks  around $\alpha\approx0.1$, which is the value we use below
(for large loop case). For long-lived loops, the distribution can be
calculated if a scaling process is assumed (see \cite{alexbook}). In the radiation era it is
\begin{eqnarray}
n(l,t)&=&\chi_r t^{-\frac{3}{2}} (l+\Gamma G \mu t)^{-\frac{5}{2}},
\nonumber
\\
&\,& l < \alpha \,t,\,\,\, t<t_{t_{eq}} \label{radiation density}
\end{eqnarray}
where $\chi_r \approx 0.4 \zeta \alpha^{1/2}$, and $\zeta$ is a
parameter related to the correlation length of the network
\cite{SCMMCR}. The numerical value of $\zeta$ is found in numerical
simulations of radiation era evolution to be about $15$ (see Table
10.1 in \cite{alexbook}). The upper bound on the
length arises because no loops are formed with sizes larger than
$\alpha t$. For $t>t_{eq}$ (the matter era) the distribution has two
components, loops formed in the matter era and survivors from the
radiation era. Loops formed in the matter era have lengths
distributed according to,
\begin{eqnarray}
n_{1}(l,t)&=&\chi_m t^{-2} (l+\Gamma G \mu t)^{-2}, \nonumber
\\
&& \alpha t_{t_{eq}} - \Gamma G \mu (t-t_{t_{eq}}) < l < \alpha
t,\,\,\, t>t_{t_{eq}} \label{matter density}
\end{eqnarray}
with $\chi_m \approx 0.12\zeta$, with $\zeta \approx 4$ (see Table
10.1 in \cite{alexbook}).  The lower bound on the
length is due to the fact that the smallest loops present in the
matter era started with a length $\alpha\, t_{eq}$ when they were
formed and their lengths have since decreased due to gravitational
wave emission. Additionally there are loops formed in the radiation
era that survive into the matter era. Their lengths are distributed
according to,
\begin{eqnarray}
n_{2}(l,t)&=&\chi_r t_{eq}^{1/2}\, t^{-2} (l+\Gamma G \mu
t)^{-\frac{5}{2}}, \nonumber
\\
&\,& l < \alpha\, t_{t_{eq}}- \Gamma G \mu (t-t_{t_{eq}}),\,\,\,
t>t_{t_{eq}}, \label{radiation survivors}
\end{eqnarray}
where the upper bound on the length comes from the fact that the
largest loops formed in the radiation era had a size $\alpha\,
t_{eq}$
but have since shrunk due to gravitational wave emission.

The cusp spectrum has been calculated in \cite{Siemens Mandic
Creighton} and the result shows that the spectrum is flat for {\it
larger}\footnote{In the following section we show that the spectrum
is flat for $f\gg \frac{H_0\sqrt{z_{eq}}}{\alpha}$ for small loops
and for $f\gg \frac{H_0\sqrt{z_{eq}}}{G\mu \Gamma}$ for large
loops.} values of $f$. Later we will show that this is also the case
for kink spectrum. This is rather unexpected since $\Omega(f)$ has
an explicit $f^{-\frac{4}{3}}$ and $f^{-\frac{1}{3}}$ dependence for
cusps and kinks, respectively. The only other $f$ dependence comes
from the $\Theta$ functions. In the following section we show
analytically that the $f$ dependence coming from the $\Theta$
function is of the form $f^{\frac{4}{3}}$ and $f^{\frac{1}{3}}$ for
cusps and kinks respectively so that the spectrum is indeed flat for
large values of the frequency $f$.

Before we start calculating the SBGW, we should mention a crucial
observation due Damour and Vilenkin \cite{DV1}. SBGW generated by a
network of cosmic strings includes bursts which occur infrequently,
and the computation of $\Omega_{\rm gw}(f)$ should not be biased by
including these large rare events (i.e. events with low rate). If
the loop density is taken of the form given in Eq. (\ref{density
small loop}), the rate is specified by the redshift only.  Therefore
the condition on the rate can be implemented by  a cutoff on
redshifts such that large events for which the rate is smaller than
the relevant time-scale of the experiment are excluded (see
Eq.~(6.17) of \cite{DV1}). However, when loops are large the
situation is more complicated because at any given redshift there
are loops of many different sizes given in Eqs. (\ref{radiation
density}) and Eq. (\ref{matter density}). This case has been dealt
with in \cite{Siemens Mandic Creighton} as follows: instead of
integrating over the variables $l$ and $z$ in Eq. (\ref{background})
one integrates over $h$ and $z$ where $h$ is defined in Eqs.
(\ref{cusp radiation in redshift}) and (\ref{kink radiation
redhift}) and imposes the cutoff limit on the $h$ integral. The
cutoff is defined as
\begin{equation}\label{rare events1}
\int _{h^*} ^\infty dh \int dz  \frac{d^2R}{dz dh} = f,
\label{cutoff in h}
\end{equation}
where $ \frac{d^2R}{dz dh}= \frac{d^2R}{dz dl}\frac{dl}{dh}$. Eq.
(\ref{cutoff in h}) is solved for $h^*$ and used to exclude  rare
event using the following integral (instead of Eq.
(\ref{background}))
\begin{equation}\label{rare events2}
\Omega_{\rm gw}(f) = \frac{4 \pi^2}{3H^2_0}f^3 \int_0 ^{h^*} dh \,
h^2 \int dz \,  \frac {d^2R}{dzdh}. \label{e:Omega(f)3}
\end{equation}
This procedure removes large amplitude events (those with strain
$h>h^*$) that occur at a rate smaller than $f$. Fig. \ref{figure
small loop}  shows the spectrum for kinks and cusps for small loops.
For the top curves (red and green) we have, $G\mu=2 \times 10^{-6}$,
$p=10^{-3}$ and $\epsilon=10^{-4}$, whereas for the bottom two
curves (blue and pink) $G\mu=10^{-7}$, $p=5\times10^{-3}$ and
$\epsilon=1$ ( $\epsilon\equiv\frac{\alpha}{\Gamma G \mu}$ ).
\begin{figure}[h!]
\begin{center}
\includegraphics[height=100 mm,angle=270]{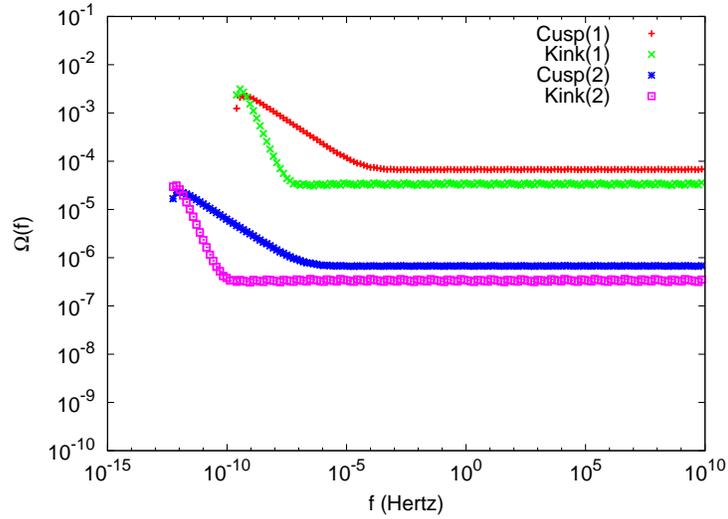}
\caption{Kink and Cusp spectrum for small loops: (1) $G\mu=2 \times
10^{-6}$, $p=10^{-3}$ and $\epsilon=10^{-4}$, (2) $G\mu=10^{-7}$,
$p=5\times10^{-3}$ and $\epsilon=1$ .}\label{figure small loop}
\end{center}
\end{figure}\\
Fig. \ref{figure large loop} shows the spectrum for large loops. For
the top curves (blue and pink), which are almost identical, we have,
$G\mu= 10^{-7}$ and $p=5\times 10^{-3}$, whereas for the bottom two
curves (red and green) $G\mu=10^{-9}$ and $p=5\times 10^{-2}$ .
\begin{figure}[h!]
\begin{center}
\includegraphics[height=100 mm,angle=270]{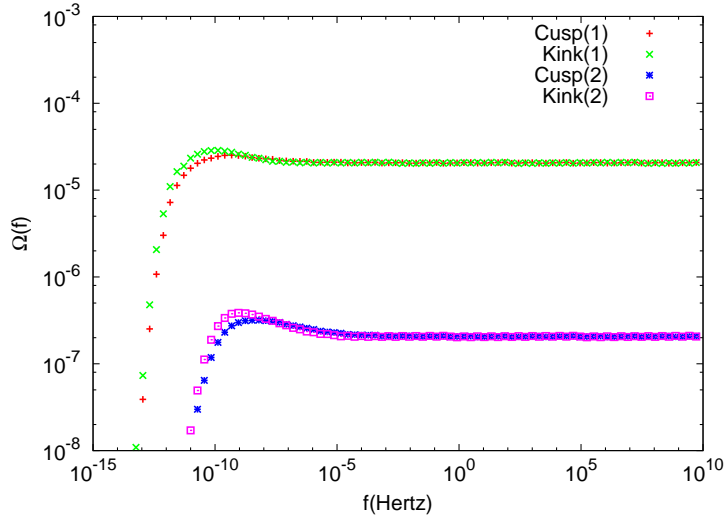}
\caption{Kink and Cusp spectrum for large loops: (1) $G\mu= 10^{-7}$
and $p=5\times 10^{-3}$, (2) $G\mu=10^{-9}$ and $p=5\times
10^{-2}$.}\label{figure large loop}
\end{center}
\end{figure}
Here we note that for $f\gg\frac{H_0}{G\mu}$, the spectrum is flat
for both cusps and kinks.\\
\section{Analytical Approximation for the Stochastic Background
}\label{Analytical app}In this section we evaluate the spectrum
analytically and show that the spectrum is constant for large values
of $f$. Our main goal is the discuss the dependence of the spectrum
on the parameters: $G\mu$, $\epsilon\equiv\frac{\alpha}{\Gamma G
\mu}$ and $p$ for small loops and $G\mu$ and $p$ for large loops. We
limit our discussion to large values of $f$, for which the spectrum
gets the dominant contribution from the loops in the radiation era.
Matter era loops contribute to lower frequency part of the
spectrum.\footnote{ It is relatively easier to verify this in the
case of small loops. If one limits the redshift integration in Eqs.
(\ref{z integral cusp radiation}) and (\ref{z integral kink
radiation}) to matter domination and uses the corresponding
approximate cosmological functions, it is found that $\Omega(f)$
depends on the negative powers of $f$, which are negligible for
large $f$. The same argument also applies to the large loop case.}
Since we want to get an estimate of the spectrum we will neglect the
complications arising from removing rare burst.
 In the radiation domination, $z>z_{eq}=\sqrt{\Omega_R}\simeq5440$, the Hubble function in Eq. (\ref{hubble}), can be approximated  as
\begin{equation}\label{hubble radiation}
    {\mathcal H}(z)\simeq\sqrt{\Omega_R}z^2=\frac{z^2}{2 \sqrt{z_{eq}}}.
\end{equation}
The cosmological functions that appear in the stochastic background
radiation formula can be approximated as
\begin{equation}\label{phit}
    \varphi_t(z)=\int_z^\infty\frac{dz'} {(1+z') {\mathcal H}(z')}\simeq\int_z^\infty\frac{dz'}
    {z' {\mathcal H}(z')}\simeq\sqrt{z_{eq}}\,z^{-2}.
\end{equation}
\begin{equation}\label{phir}
    \varphi_r(z)=\int_0^z\frac{dz'} { {\mathcal H}(z')}=\int_0^{z_{eq}}\frac{dz'}
    { {\mathcal H}(z')}+\int_{z_{eq}}^z\frac{dz'} { {\mathcal H}(z')}\simeq \,3.6.
\end{equation}
\begin{equation}\label{phiV}
    \varphi_V(z)=\frac{4 \pi  \varphi_r^2}{(1+z)^3  {\mathcal H}(z)}\simeq 325\,
    \sqrt{z_{eq}} \,z^{-5}.
\end{equation}
We first consider the small loop case, for which the expression for
SBGW reduces to an integral over redshift given in Eqs. (\ref{z
small loop cusp}) and (\ref{z small loop kink}). Inserting the
result in Eqs. (\ref{phit}-\ref{phiV}) into Eq. (\ref{z small loop
cusp}) we get
\begin{eqnarray}\label{z integral cusp radiation}
\Omega_{gw,R}(f)&\propto&\frac{ G\mu} {p \alpha^{1/3}
f^{1/3}}\int_{z_{eq}}^{z_{max}}\frac{dz}{z^{2/3}}\Theta\left(
1-\left[\frac{f z_{eq}^{1/2} \alpha}{H_0
z}\right]^{-1/3}\right)\propto\frac{G\mu } {p},
\end{eqnarray}
where we dropped a term with $1/f$ dependence since it is small in
large $f$ limit, and the subscript $R$ reminds us that this is the
contribution from radiation era loops. The upper limit of the
integration, $z_{max}$, is the redshift at the time of the creation
of the strings, which depends on the energy scale of the phase
transition. The result in Eq. (\ref{z integral cusp radiation}) is
valid for $\frac{ z_{eq}^{1/2}}{\alpha }\ll\frac{f}{H_0}<\frac{
z_{max}}{\alpha z_{eq}^{1/2}}$, for which the upper limit of the
integral is set by the $\Theta$-function. If $\frac{f}{H_0}>\frac{
z_{max}}{\alpha z_{eq}^{1/2}}$, the integral does not depend on $f$
and the frequency dependence of $ \Omega_{gw,R}(f)$ is given by the
prefactor, which has $f^{-1/3}$ behavior.  For kinks we get
\begin{eqnarray}\label{z integral kink radiation}
\Omega^K_{gw,R}(f)&\propto&\frac{ G\mu } {p \alpha^{2/3}
f^{2/3}}\int_{z_{eq}}^{z_{max}}\frac{dz}{z^{1/3}}\Theta\left(
1-\left[\frac{f z_{eq}^{1/2} \alpha}{H_0
z}\right]^{-1/3}\right)\propto\frac{G\mu } {p}.
\end{eqnarray}
Eqs. (\ref{z integral cusp radiation}) and (\ref{z integral kink
radiation}) show that for $\frac{ z_{eq}^{1/2}}{\alpha
}\ll\frac{f}{H_0}$ the spectrum is constant and it scales with
$G\mu/p$. The amplitude does not depend on the parameter $\alpha$,
however the spectrum shifts to the right linearly in  $\alpha$.\\
This result is in perfect agreement with Fig. \ref{figure small
loop}. For the bottom curves $\frac{G\mu}{p}=2 \times 10^{-5}$ where
as $\frac{G\mu}{p}=2 \times 10^{-3}$ for the top curves, which have
two orders of magnitude larger amplitude, exactly agreeing with the
figure. Furthermore,  the top curves ($\epsilon=10^{-4}$) are
shifted to the right compared to the bottom curves ($\epsilon=1$) by
about $4$-orders in $f$ as predicted by our results above.

 Now we consider large loops in the radiation domination, for which the density $n(l,t)$ is given in Eq. (\ref{radiation
density}), where  $t$ is to be replaced with $\varphi_t(z)/H_0$.
Substituting  the results in Eqs. (\ref{phit}-\ref{phiV}) into Eq.
(\ref{background}) we get
\begin{eqnarray}\label{domega}
\Omega_{gw,R}(f)&=&A(f)\int dz\int dl \frac{z (l
z)^{-\frac{1}{3}}}{(l z^2 + \beta
\delta)^{\frac{5}{2}}}\Theta(1-\frac{1}{f z
   l})\Theta(\frac{\beta}{z^2}-l)\nonumber\\
  &=& A(f)\int_{z_{eq}}^{z^*}dz\int_{\frac{1}{f}}^{\frac{\beta}{z}} du
   \frac{u^{-\frac{1}{3}}}{(u z + \beta \delta)^{\frac{5}{2}}}
\end{eqnarray}
where we define
\begin{eqnarray}\label{rate}
 A(f)&=& \frac{165\, c\, \alpha^2\, \delta^2
  \chi_R }{  p \,{z_{eq}}^{1/4}H_0^{\frac{3}{2}}\Gamma^2f^{\frac{1}{3}}},
\end{eqnarray}
with $\delta=\frac{G \mu \Gamma}{\alpha}$ and
$\beta=\frac{\alpha\sqrt{z_{eq}}}{H_0}$ ($\alpha\approx0.1$ for
large loop case) and the dummy integration variable $u=l\, z$. The
upper limit of the $z$ integral, $z^*$ will be set by requiring
$\beta/z>1/f$, that is, $z<f\beta$. If $f<z_{max}/\beta$ we have,
\begin{eqnarray}\label{omega int 2}
  \Omega(f)&=&A(f)\int_{z_{eq}}^{\beta/f}dz\int_{\frac{1}{f}}^{\frac{\beta}{z}}du\frac{u^{-\frac{1}{3}}}{(u
z + \beta \delta)^{\frac{5}{2}}}
=A(f)\int_{\frac{1}{f}}^{\frac{\beta}{z_{eq}}}du\int_{z_{eq}}^{\beta/u}dz\frac{u^{-\frac{1}{3}}}{(u
z + \beta \delta)^{\frac{5}{2}}}\nonumber\\
&=&-\frac{2}{3}A(f)\int_{\frac{1}{f}}^{\frac{\beta}{z_{eq}}}\frac{du}{u^{\frac{4}{3}}}\left(\frac{1}{(\beta+\beta
\delta)^{\frac{3}{2}}}-\frac{1}{(u z_{eq}+\beta
\delta)^{\frac{3}{2}}} \right).
\end{eqnarray}
If $\frac{1}{f}<\frac{\delta\beta}{z_{eq}}=\frac{G\mu \Gamma}{H_0
\sqrt{z_{eq}}}$, we can split the integration range
$[1/f,\beta/z_{eq}]$ in the second integral into
$[1/f,\delta\beta/z_{eq}]$ and $[\delta\beta/z_{eq},\beta/z_{eq}]$
and neglect $u z_{eq}$ and $\beta \delta$ respectively in these two
integrals.
 Combining all terms and keeping the lowest order in $\delta$ we
get,
\begin{eqnarray}\label{omega final large f}
  \Omega_{gw,R}(f)&=&A(f)\left(\frac{2 f^{\frac{1}{3}}}{ (\delta\beta)^{\frac{3}{2}}}-
  \frac{18 {z_{eq}}^{\frac{1}{3}}}{11
  (\delta\beta)^{11/6}}\right)=\frac{330\, c\, \alpha^2\, \delta^{\frac{1}{2}}
  \chi_R }{  p \,{z_{eq}}^{1/4}H_0^{\frac{3}{2}}\Gamma^2\beta^{\frac{3}{2}}}
  \nonumber\\
  &\simeq&3.2 \times10^{-4} \frac{\sqrt{G \mu}}{
  p},  f>\frac{3.6 \times 10^{-18}}{G\mu}Hz
  \end{eqnarray}
The calculation for the case of kink is very similar to cusp case,
following the same steps we get
\begin{equation}\label{omega final large f KINK}
    \Omega_{gw,R}^{K}(f)\simeq 3.2 \times10^{-4} \frac{\sqrt{G \mu}}{
  p},\,\, f>\frac{3.6 \times 10^{-18}}{G\mu} Hz
\end{equation}
which is identical to the cusp result. Eqs. (\ref{omega final large
f}) and (\ref{omega final large f KINK}) show that the distribution
is flat for $f>\frac{3.6 \times 10^{-18}}{G\mu} Hz$ and its
amplitude scales with $\sqrt{G \mu}/p$, which is in excellent
agreement with Fig. \ref{figure large loop}. The flat value of the
spectrum for the top curves ($G\mu=10^{-7}$ and $p=5\times10^{-3}$)
is $2.1\times 10^{-5}$ and  for the bottom curve ($G\mu=10^{-9}$ and
$p=5\times10^{-2}$) is $2.1\times 10^{-7}$. These results are to be
compared with the analytical results $2.0\times 10^{-5}$ and
$2.0\times 10^{-7}$ predicted by Eqs. (\ref{omega final large f})
and (\ref{omega final large f KINK}).\\
It is important to note that, in this paper we assume that the
number of kinks, $N$, is order of one. This assumption enters in the
estimation $|\ddot X_\pm|\sim \frac{2 \pi}{l}$, and if there are $N$
kinks on strings, it needs to be replaced by $|\ddot X_\pm|\sim
\frac{2 \pi}{l/N}$.  The replacement of $l$ with $l/N$ should also
be done in the opening angle of the cone of the radiation, Eq.
(\ref{theta max}),which will result in a nontrivial dependence on
$N$. However we can simply convert the resultant expression to the
one we calculated in Eq. (\ref{z small loop kink} ) by defining
$\alpha=\alpha' N$. Since we have shown that $\alpha$ has the effect
of moving the spectrum horizontally, one effect of having $N$ kinks
will be shifted spectrum compared to one kink spectrum. The other
effect will be an overall scaling of the spectrum by $1/N$.
 \section{Parameter Space Constraints and Results}\label{parameter scan}
 In this section we discuss certain experimental bounds on SBGW. For
 the case of large loops the parameters are $G\mu$ and $p$, and for small loops the parameters are
 $G\mu$, $\epsilon$ and $p$. It is
 important to note that the nontrivial dependence on $p$ follows
 from excluding rare bursts as described in Eqs. (\ref{rare
 events1}) and (\ref{rare events2}) (if rare events
 were included $\Omega(f)$ would simply scale with $1/p$.)
\begin{figure}[hbtp]
$\begin{array}{cc}\label{parameter scan plots}
\includegraphics[width=2.5in]{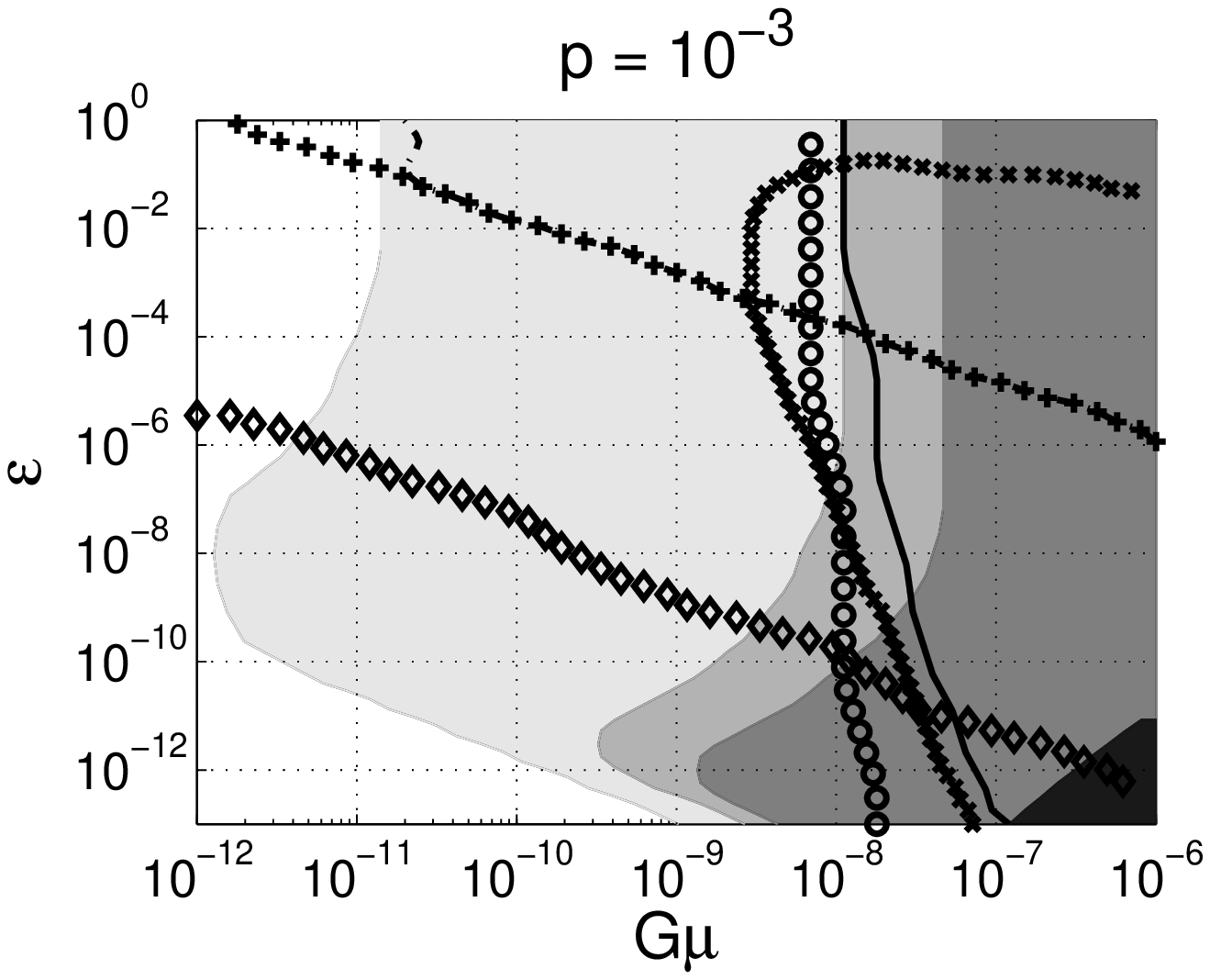} &
\hspace{1cm}
\includegraphics[width=2.5in]{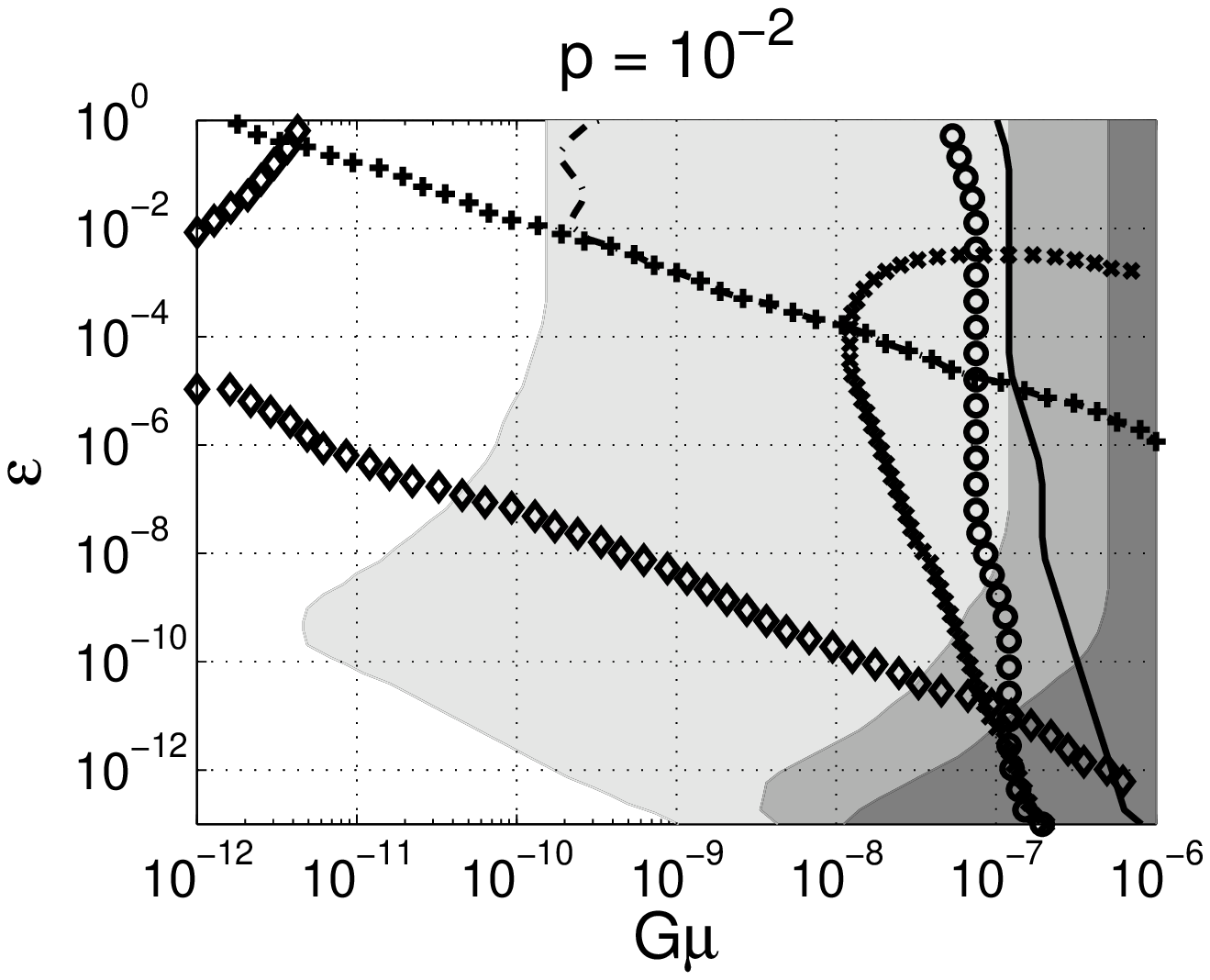} \\
\includegraphics[width=2.5in]{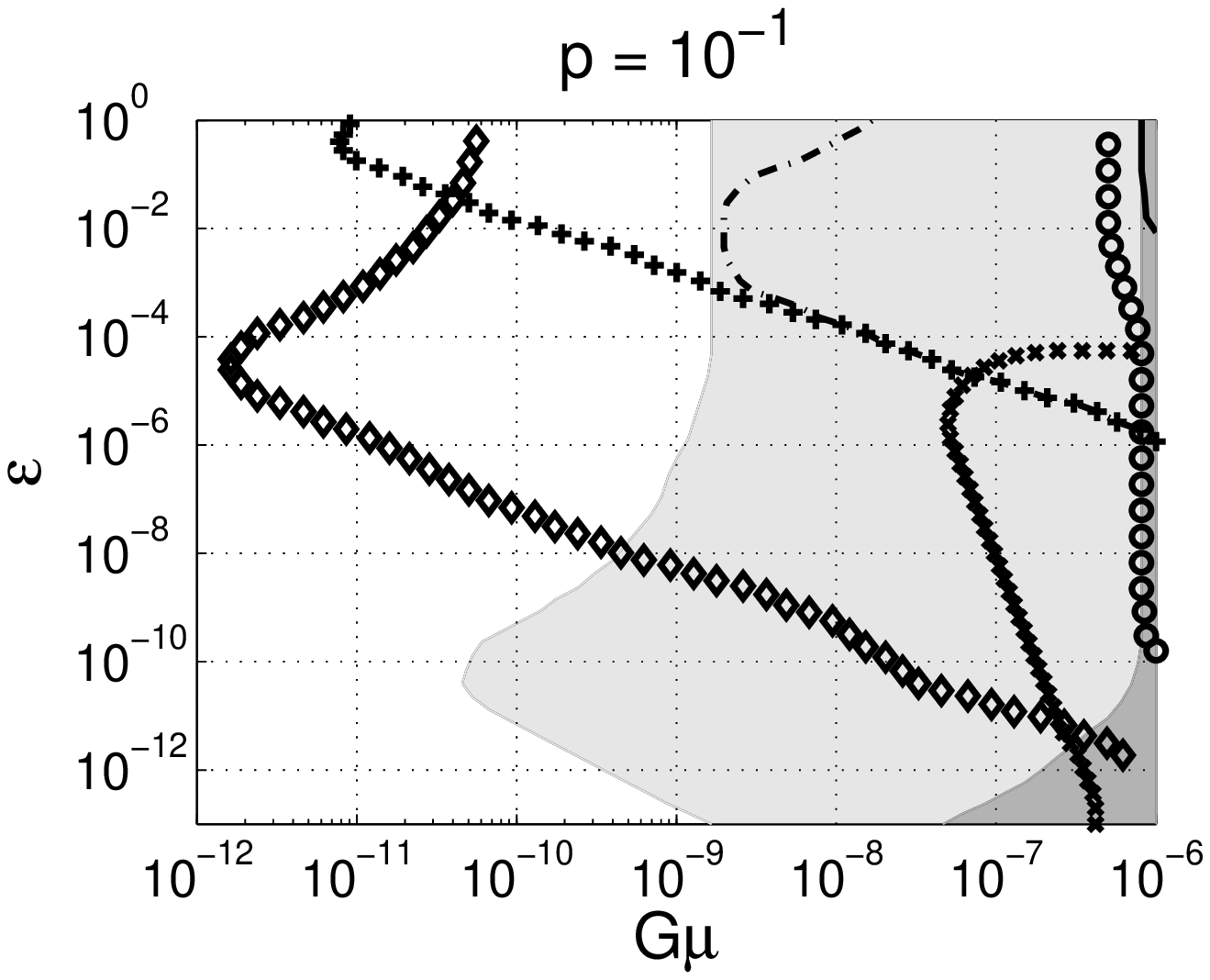} &
\hspace{1cm}
\includegraphics[width=2.5in]{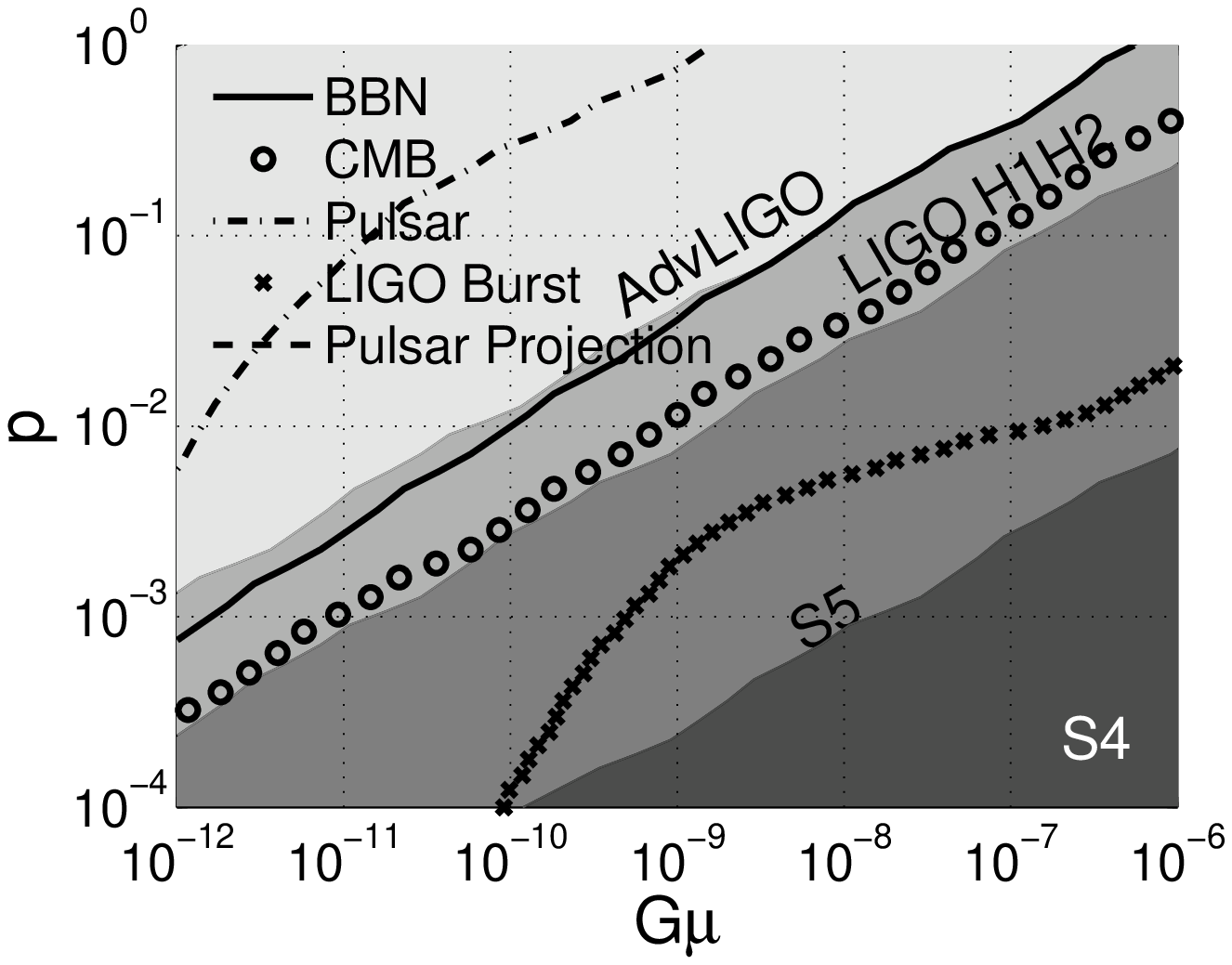} \\
\end{array}$
\caption{Top-left: Accessible regions in the $\varepsilon-G\mu$
plane
  for $p = 10^{-3}$ for small loops (loop sizes are determined by gravitational
  back-reaction). Top-right:
  Same as above for $p=10^{-2}$.  Bottom-left: Same as above for
  $p=10^{-1}$.  Bottom-right: Accessible regions in the $p-G\mu$ plane
  for the large long-lived loop models. The accessible regions are to
  the right of the corresponding curves. All models are within reach
  of LISA and advanced LIGO, and most are within the projected pulsar bound.}
\label{results}
\end{figure}\\
 Accessible regions corresponding to different
experiments and bounds are shown in Fig. 3. The shaded regions, from
darkest to lightest, are: LIGO S4 \cite{LIGOS4} limit,
  LIGO S5 \cite{S5}, LIGO H1H2 projected sensitivity (cross-correlating the data
  from the two LIGO interferometers at Hanford, WA (H1 and H2)), and
  AdvLIGO H1H2 projected sensitivity.  All projections assume 1 year
  of exposure and either LIGO design sensitivity or Advanced LIGO
  sensitivity tuned for binary neutron star inspiral search.  The
  solid black curve corresponds to the BBN \cite{BBN} bound, the dot-dashed curve
  to the pulsar bound\cite{pulsar}, the $+$s to the projected pulsar sensitivity,
  the  circles to the bound based on the CMB and matter spectra \cite{CMB},
  the $\times$s to the projected sensitivity of the LIGO burst \cite{SCMMCR} search,
  and the $\diamond$-curve  to the LISA projected sensitivity \cite{LISA}. The BBN and CMB
  bounds are integral bounds, i.e. they are upper limits for the integral of $\Omega(f)$ over $\ln f$, therefore
  a model is excluded if it predicts an integral larger than the limit. On the other hand,
  the pulsar and LIGO bounds apply in specific frequency bands, thus
  a model is excluded if it has $\Omega(f)$ larger than the limit (or projected
  sensitivity) for any $f$ in the range of the pulsar or LIGO
  experiments. The range of the redshift integral in Eq. (\ref{background}) must
  chosen properly for a given experiment. For BBN bound, the
  integration is performed for $z > 5.5 \times 10^9$.  Similarly, for the bound
based on the CMB and matter spectra, the integration is performed
for $z > 1100$. First, we  note that smaller values of $p$ are more
accessible, which  follows from the fact that the loop density is
inversely proportional to $p$. This makes cosmic superstrings more
accessible than field theoretical strings. Second, we note that LIGO
stochastic search constrains large $G\mu$, small $\epsilon$ part of
the parameter space, whereas pulsar limit constrains large $G\mu$
and large $\epsilon$ part of the parameter space. Similarly, the
LIGO burst bound applies to large $G\mu$ and intermediate $\epsilon$
part of the parameter space. Therefore large $G\mu$ part of the
parameter space is covered by these three experiments. Furthermore
since they also overlap for large $G\mu$ and intermediate
$\epsilon$, in the case of detection, the two LIGO searches could
potentially confirm each other.  We also see that the BBN and CMB
bounds are not very sensitive to $\epsilon$: the corresponding
curves are rather vertical in $\epsilon-G\mu$ plane. This result is
in perfect agreement our results  (Eqs. (\ref{z integral cusp
radiation}) and (\ref{z integral kink radiation})) that show
$\Omega(f)\propto G\mu/p$, which does not depend on $\epsilon$. For
the case of large loops, GW background is significantly larger than
the small loop one, see Figs. \ref{figure small loop} and
\ref{figure large loop}. Therefore more of the parameter space is
accessible to the current and proposed experiments, as depicted in
the right bottom panel of Fig. 3. The strongest constraint  is the
pulsar bound, which rules out cosmic (super)string models with
$G\mu>10^{-12}$ and $p< 8\times 10^{-3}$. This bound also rules out
field theoretical strings ($p=1$) with $G\mu>2\times 10^{-9}.$ One
can compare these results with the case where only cusps are
included \cite{Siemens Mandic Creighton}. In that case cosmic
(super)string models with $G\mu>10^{-12}$ and $p< 3\times 10^{-3}$
and field theoretical strings  with $G\mu>10^{-9}$ are ruled out.
This result illustrates that kinks contribute to SBGW at the same
order as cusps.
\section{Acknowledgment} We would like to thank Marco Peloso for
useful discussions. S.\"O. is supported by the Graduate School at
the University of Minnesota under the Doctoral Dissertation
Fellowship, X. S. is supported in part by NSF Grant No. PHY-0758155
and the Research Growth Initiative at the University of
Wisconsin-Milwaukee and V. M. is supported in part by NSF Grant No.
PHY0758036.


\begin{thebibliography} {99}
\bibitem{ANO} H. B. Nielsen and P. Olesen, Nuc. Phys. {\bf B61}, 45 (1973)
\bibitem{kibble76} T.W.B. Kibble, J. Phys. {\bf A9}  1387 (1976).
\bibitem{alexbook}
A.~Vilenkin and E.~Shellard, Cosmic strings and other Topological Defects
  (Cambridge University Press, 2000).
\bibitem{Jeannerot:2003qv}
R.~Jeannerot, J.~Rocher, and M.~Sakellariadou, Phys. Rev.
\textbf{D68} 10 3514 (2003).
\bibitem{Zeldovich}
Y.~B.~Zeldovich, M.N.R.A.S. {\bf 192}, 663 (1980).
\bibitem{AV81a}
A.~Vilenkin,
Phys.\ Rev.\ Lett.\  {\bf 46}, 1169 (1981) [Erratum-ibid.\  {\bf
46}, 1496 (1981)].
\bibitem{book anderson}M. R. Anderson,
{\it The Mathematical Theory of cosmic strings}, Institute of
Physics Publishing, 2003. V.A.Gasilov, V.I.Maslyankin and
\bibitem{khlopov1} M.Yu.Khlopov, Astrofizika
, {\bf V.23}, PP.191-201 (1985). [English translation: Astrophysics,
{\bf V.23}, NO.1/JAN, PP. 485-491 (1986)]
\bibitem{khlopov2}
M.V.Sazhin and M.Yu. Khlopov  Astron. Zh. , {\bf V. 66}, PP. 191-193
(1989). [English translation: Sov. Astron. , {\bf V.33}, no.1, P. 98
(1989)]
\bibitem{cstrings} N. Jones et al., JHEP {\bf 0207}
051 (2002); S. Sarangi, S.H.Henry Tye, Phys.Lett. {\bf B536} 185
(2002); G. Dvali, A. Vilenkin, JCAP {\bf0403}  010 (2004); N. Jones
et al., Phys.Lett.  {\bf B563}  6 (2003); E.J. Copeland et al., JHEP
{\bf 0406}  013 (2004).
 \bibitem{Nitta} M. Eto, K. Hashimoto, G. Marmorini, M. Nitta, K. Ohashi and W.
 Vinci JCAP {\bf 0509}  004 (2005) [arXiv: hep-th/0506022v2].
\bibitem{dvali vilenkin 2004} G. Dvali, A. Vilenkin, JCAP {\bf 0403}  010 (2004).
\bibitem{jjp} M.G. Jackson, N.T. Jones and J. Polchinski, JHEP {\bf 0510}  013 (2005).
\bibitem{Leblond:2009fq}  L.~Leblond, B.~Shlaer and X.~Siemens,
  Phys.\ Rev.\   {\bf D79}, 123519 (2009),
  [arXiv: astro-ph/0903.4686].
  \bibitem{stoica tye} N. Jones, H. Stoica, S.H.Henry Tye, Phys.Lett.
{\bf B563}  6 (2003).
\bibitem{Sakellariadou 2005} M. Sakellariadou, JCAP {\bf 0504}  003 (2005).
\bibitem{avgoustidis shellard} A. Avgoustidis and E.P.S Shellard, [arXiv: astro-ph/0512582].
\bibitem{CMB1}L.~Pogosian, M.~C.~Wyman and I.~Wasserman, [arXiv: astro-ph/0403268].
\bibitem{CMB2} E.~Jeong and G.~F.~Smoot, [arXiv: astro-ph/0406432].
\bibitem{DV1}
T.~Damour and A.~Vilenkin,
Phys.\ Rev.\ Lett.\  {\bf 85}, 3761 (2000), [arXiv: gr-qc/0004075].
\bibitem{DV2}
T.~Damour and A.~Vilenkin,
Phys.\ Rev.\  {\bf D64}, 064008 (2001), [arXiv: gr-qc/0104026].
\bibitem{SCMMCR} X. Siemens et al., Phys. Rev {\bf D73}  105001 (2006).
\bibitem{pol1} J. Polchinski, [arXiv: hep-th/0410082]; J. Polchinski, [arXiv: hep-th/0412244].
\bibitem{Abbott:2009rr}
  B.~P.~Abbott {\it et al.}  [LIGO Scientific Collaboration],
  Phys.\ Rev.\   {\bf D80}, 062002 (2009), [arXiv: astro-ph/0904.4718 ].
\bibitem{Siemens Mandic Creighton} X. Siemens, V. Mandic and J. Creighton, Phys.\ Rev.\ Lett.{\bf  98}, 111101 (2007).
\bibitem{Kawasaki} M. Kawasaki, K. Miyamoto and K. Nakayama,
[arXiv: astro-ph/1002.0652].
\bibitem{Kawasaki2} M. Kawasaki, K. Miyamoto and K. Nakayama,
[arXiv: astro-ph/1003.3701].
\bibitem{Weinberg} S.~Weinberg, {\it Gravitation and Cosmology} Wiley, New York, 1972.
\bibitem {copeland kibble}E. J. Copeland and T. W. B. Kibble, Phys. Rev. {bf D80}, 123523
(2009), [arXiv: astro-ph/0909.1960].
\bibitem{RSB} C. Ringeval, M. Sakellaridou and F. Bouchet, [arXiv: astro-ph/0511646].
\bibitem{shellardrecentsim}  C.J.A.P. Martins and E.P.S. Shellard,
[arXiv: astro-ph/0511792].
\bibitem{VOV} V. Vanchurin, K.D. Olum and A. Vilenkin, [arXiv: gr-qc/0511159].
\bibitem{LIGOS4}
Abbott B. et al.,  Astrophys. J. {\bf 659}, 918-930 (2007).
\bibitem{S5} B. P. Abbott et al. (LIGO and Virgo Collaborations),
Nature {\bf 460}, 990 (2009).
\bibitem{BBN}
Cyburt, R.H. et al, Astropart. Phys. {\bf 23}, 313-323 (2005).
\bibitem{pulsar}
Jenet, F.A. et al., Astrophys. J. {\bf 653}, 1571-1576 (2006)
\bibitem{CMB} Smith, T.L., Pierpaoli, E., and Kamionkowski, M.,
Phys. Rev. Lett. {\bf 97}, 021301 (2006).
\bibitem{LISA} P.~L. Bender, K. Danzmann, and the LISA Study Team, MPQ233 (1998).




\end{thebibliography}
\end{document}